\documentclass[twocolumn,showpacs,preprintnumbers,amsmath,amssymb]{revtex4}
\usepackage{graphicx}
\usepackage{dcolumn}
\usepackage{bm}
\begin{document}

\title{ Protecting quantum correlations of two qubits in independent non-Markovian
environments by bang-bang pulses}

\author{Hang-Shi Xu}
\author{Jing-bo Xu}
 \email{xujb@zju.edu.cn}
\affiliation{Zhejiang Institute of Modern Physics and Physics Department, Zhejiang University, Hangzhou 310027,
People's Republic of China}

\begin{abstract}
We investigate how to protect quantum correlations for two qubits each locally interacting with its own non-Markovian environment by making use of bang-bang pulses. It is shown that the quantum discord dynamics presents the phenomenon of
sudden change for some certain initial states. We also find that the amount of quantum correlation between two qubits can be improved by applying a train of pulses and protected more effectively with shorter interval pulses or longer reservoir correlation time.
\end{abstract}


\maketitle

\section{Introduction}
Quantum correlations has become an hot topic in quantum information and quantum computation and attracted much attention in recent years [1-5]. Entanglement is a special kind of quantum correlation and has been widely used in quantum information processing [1]. However, it has been reorganized that the quantum discord which is firstly introduced by Ollivier and Zurek [2], has an advantage over entanglement and can be responsible for the computational speedup for certain quantum tasks [3-6]. Quantum discord, which is defined as the difference between the quantum mutual information and the classical correlation, is nonzero even for separate mixed states and may be regarded as a more general and fundamental resource in the quantum information processing.

Recently, the dynamics of entanglement and quantum discord for some open systems has attracted
much attention [7-10]. In the non-Markovian environments, the memory stored in the environment has been considered, some of the initial quantum correlation which is lost during the dissipative dynamics can return to the qubits [7]. It has been shown that the discord is more robust than entanglement under the Markovian
environments and vanishes only at some time points under non-Markovian environment [7, 8].  Furthermore, quantum discord has been predicted to show the sudden transition between classical and quantum decoherence [9] and the phenomenon of sudden change [10] for some certain initial states. Some of these phenomena have been verified by the recent experiments [10,11]. The enhancement of quantum discord for the system of cavity QED without dissipation by bang-bang pulses has been investigated [12]. Recently experimental suppressions of decoherence in a ring cavity and solid-state qubits using bang-bang decoupling technique have also been reported [13].

In this paper, we propose a scheme of protecting quantum correlations for two qubits each locally interacting with its own non-Markovian reservoir by means of bang-bang pulses. Using the master equation approach, we obtain an explicit expression for the reduced density matrix of the system. Firstly, we assume that the two qubits are initially prepared in the states with maximally mixed marginals [14] and show that the quantum discord of two qubits displays the phenomenon of sudden change for some certain initial states. Furthermore, we investigate how the bang-bang pulses affect the quantum discord and entanglement of two qubits and the phenomenon of sudden change. It is found that the bang-bang pulses can be used to improve the quantum correlations of two qubits. The pulses with shorter interval can protect the quantum correlations of two qubits more effectively. The influence of reservoir correlation time and relaxation time scale are also investigated.

\section{Dynamics of two qubits in independent non-Markovian environments  with bang-bang pulses}
In this section we investigate the dynamics of quantum correlations for two qubits
each locally interacting with its own non-Markovian reservoirs with bang-bang pulses. The Hamiltonian
of one qubit interacting with its own reservoir with bang-bang pulses is given by
\begin{equation}
H=H_0+H_I+H_P(t)
\end{equation}
with
\begin{equation}
H_0=\frac{\omega_0}{2}\sigma_z+\sum_k\omega_k a_k^\dag a_k; \quad
H_I=\sum_kg_k(\sigma_+a_k+\sigma_-a_k^\dag),
\end{equation}
where $a_k$ and $a_k^\dag$ denote the annihilation and creation
operators of the reservoir field and $\sigma_z=|e\rangle\langle
e|-|g\rangle\langle g|$, $\sigma_+=|e\rangle\langle g|$,
$\sigma_-=|g\rangle\langle e|$ are the atomic operators. The index $k$ labels different
field modes of the reservoir with coupling constant $g_k$ and frequencies $\omega_k$.
$H_P$ is the Hamiltonian for a train of identical pulses of duration $\tau$, i.e.,
\begin{equation}
H_P(t)=V\sigma_z\sum_{n=0}^\infty \theta(t-T-n(T+\tau))\theta((n+1)(T+\tau)-t),
\end{equation}
where  $T$ is the time interval between two consecutive pulses and the amplitude $V$ of the control
field is specified to be $\frac{\pi}{2\tau}$, which means that we consider the $\pi$-pulse only.
For the case that the pulses is strong enough, i.e. the duration $\tau\rightarrow 0$, the time evolution operator with pulse presented is given by
$U_{P}=\exp[-i(H_0+H_I+ \frac{\pi}{2\tau}\sigma_z)\tau]\simeq\exp[- i\frac{\pi}{2}\sigma_z]$.

By making use of the master equation approach for the reduced density matrix, to the
second order in perturbation theory (Born approximation), we obtain the following equation
for the reduced density matrix of the qubit [15]
\begin{equation}
\frac{d\rho(t)}{dt}=-\int_0^tdt'Tr_R[H_I(t),[H_I(t'),\rho(t')\otimes \rho_R]],
\end{equation}
where $\rho(t)$, $\rho_R$ are the density matrix of the qubit and reservoir respectively, and $H_I(t)$ is the
Hamiltonian in the interaction picture which given by
\begin{equation}
H_I(t)=U^\dag(t) H_IU(t),
\end{equation}
with $U(t)=\widehat{T}e^{-i\int_0^tdt'[H_0+H_P(t')]}$. Using Eq. (3), we can obtain $U(t)$ and $H_I(t)$ as follows,
\begin{equation}
U(t)=[U_PU_0(T)]^nU_0(t-nT),
\end{equation}
\begin{equation}
H_I(t)=\sum_kg_k(-1)^{[\frac{t}{T}]}(e^{i\omega_0t}\sigma_+a_k+e^{-i\omega_0t}\sigma_-a_k^\dag).
\end{equation}
where $n=[\frac{t}{T}]$, $[\quad]$ denotes the integer part, and $U_0(t)=e^{-iH_0t}$.

Assuming that the qubit is in resonance with the cavity mode, the reservoir spectral density is the Lorentzian
spectral distribution [16]:
\begin{equation}
J(\omega)=\frac{1}{2\pi}\frac{\gamma_0\lambda^2}{(\omega_0-\omega)^2+\lambda^2},
\end{equation}
where $\gamma_0$ is the Markovian decay rate connected to the relaxation time scale $\tau_R(\simeq \gamma_0^{-1})$ and $\lambda$ is the spectral width of the coupling connected to the reservoir correlation time $\tau_B(\simeq \lambda^{-1})$. The Markovian and non-Markovian regimes are distinguished by $\gamma_0$ and $\lambda$: $\gamma_0< \lambda/2$ means
the Markovian regime and $\gamma_0 > \lambda/2$ corresponds to the non-
Markovian regime [16]. Using Eq. (4), (7) and (8), the master equation becomes
\begin{equation}
\frac{d\rho(t)}{dt}=\int_0^tdt'K(t,t')\mathcal{L}\rho(t'),
\end{equation}
\begin{equation}
\mathcal{L}\rho(t)=\sigma_-\rho\sigma_+-\frac{1}{2}\sigma_+\sigma_-\rho-\frac{1}{2}\rho\sigma_+\sigma_-,
\end{equation}
where $\mathcal{L}$ is the Markovian Liouvillian and $K(t,t')$ is the memory kernel which given by,
\begin{equation}
K(t,t')=(-1)^{[\frac{t}{T}]+[\frac{t'}{T}]}k(t'-t),
\end{equation}
where $k(t)=\int d\omega J(\omega)e^{i(\omega-\omega_0)t}=\frac{\gamma_0\lambda}{2}e^{-\lambda t}$ is the correlation function.
The factor $(-1)^{[\frac{t}{T}]+[\frac{t'}{T}]}$ is induced by the bang-bang pulses, which averaging to zero the unwanted
interaction with the environment when $T\rightarrow 0$.

The dynamics of a qubit $S$ can be described
by the reduced density matrix [15, 16]
\begin{eqnarray}
\rho^S(t)=
\left(
\begin{array}{c}
\rho^S_{11}(0)P_t  \quad \rho^S_{10}(0)\sqrt{P_t}\\
\rho^S_{01}(0)\sqrt{P_t} \quad 1-\rho^S_{11}(0)P_t
\end{array}\right),
\end{eqnarray}
where the function $P_t$ obeys the differential equation
\begin{equation}
\frac{dP_t}{dt}=\int_0^tdt'K(t,t')P_{t'}.
\end{equation}

The general solution of $P_t$ can be derived as,
\begin{equation}
P_t=e^{-\lambda t}[A_n\cos\frac{d(t-nT)}{2}+B_n\sin\frac{d(t-nT)}{2}]^2,
\end{equation}
where $t\in[nT, (n+1)T]$ and $d=\sqrt{2\gamma_0\lambda-\lambda^2}$. Using the initial values $A_0=1$ and $B_0=\frac{\lambda}{d}$ and boundary conditions, we can obtain recurrence relations of the constant coefficients $A_n$ and $B_n(n>0)$,
$$A_n=e^{-\frac{\lambda}{2}T}[A_{n-1}\cos\frac{dT}{2}+B_{n-1}\sin\frac{dT)}{2}],$$
$$B_n=e^{-\frac{\lambda}{2}T}[A_{n-1}\cos\frac{dT}{2}+B_{n-1}\sin\frac{dT)}{2}]$$
\begin{equation}
-\frac{2\lambda}{d}e^{-\frac{\lambda}{2}T}[A_{n-1}\sin\frac{dT}{2}+B_{n-1}\cos\frac{dT)}{2}].
\end{equation}

In the standard product basis $\{|1\rangle\equiv|11\rangle, |2\rangle\equiv|10\rangle, |3\rangle\equiv|01\rangle, |4\rangle\equiv|00\rangle \}$, the elements for the reduced
density matrix $\rho^{AB}(t)$ for the two-qubit system [16] can be constructed as,
$$\rho^{AB}_{11}(t)=\rho^{AB}_{11}(0)P_t^2,$$
$$\rho^{AB}_{22}(t)=\rho^{AB}_{22}(0)P_t+\rho^{AB}_{11}(0)P_t(1-P_t),$$
$$\rho^{AB}_{33}(t)=\rho^{AB}_{33}(0)P_t+\rho^{AB}_{11}(0)P_t(1-P_t),$$
\begin{equation}
\rho^{AB}_{44}(t)=1-[\rho^{AB}_{11}(t)+\rho^{AB}_{22}(t)+\rho^{AB}_{33}(t)],
\end{equation}
$$\rho^{AB}_{12}(t)=\rho^{AB}_{12}(0)P_t^{\frac{3}{2}}, \rho^{AB}_{13}(t)=\rho^{AB}_{13}(0)P_t^{\frac{3}{2}},$$
$$\rho^{AB}_{14}(t)=\rho^{AB}_{14}(0)P_t, \rho^{AB}_{23}(t)=\rho^{AB}_{23}(0)P_t$$
$$\rho^{AB}_{24}(t)=\sqrt{P_t}[\rho^{AB}_{24}(0)P_t+\rho^{AB}_{13}(0)(1-P_t)],$$
\begin{equation}
\rho^{AB}_{34}(t)=\sqrt{P_t}[\rho^{AB}_{34}(0)P_t+\rho^{AB}_{12}(0)(1-P_t)],
\end{equation}
and $\rho^{AB}_{ij}(t)=\rho^{AB*}_{ji}(t)$.

\section{Protecting quantum correlations of two qubits in independent non-Markovian environments by bang-bang pulses}
In this section we investigate how to protect quantum discord and entanglement for two qubits
each locally interacting with its own non-Markovian reservoirs with bang-bang pulses. The definition of quantum discord is based on quantum mutual information which contains both classical and
quantum correlations. For a bipartite system $\rho^{AB}$, its total correlations can be measured by their quantum
mutual information [2, 17]
\begin{equation}
I(\rho^{AB})=S(\rho^{A}) + S(\rho^{B}) -
S(\rho^{AB}),
\end{equation}
where $S(\rho)=-Tr(\rho\log_2\rho)$ is the von Neumann entropy, and $\rho^{A}$ and $\rho^{B}$ denote the reduced
density matrices of parts $A$ and $B$, respectively. The quantum discord is defined
as the difference between the quantum mutual information and the
classical correlation [2, 17],
\begin{equation}
Q(\rho^{AB}) =I(\rho^{AB}) -
C(\rho^{AB}),
\end{equation}
where $C(\rho^{AB})$ is the classical correlation which depends on the maximal information obtained with
measurement on one of the subsystems and can be expressed as
\begin{equation}
C(\rho^{AB})=\max_{\{B_{k}\}}[S(\rho^{A}) -
S(\rho^{AB}|{\{B_{k}\})}],
\end{equation}
where $\{B_{k}\}$ is a complete set of projectors preformed on
subsystem $B$ locally, $S(\rho^{AB}|\{B_{k}\})=\sum_k p_kS(\rho_{k})$
is the quantum conditional entropy,
$\rho_k=1/p_k (I\otimes B_k) \rho^{AB}
(I\otimes B_k)$ is the conditional density operator and
$p_k=\mathrm{tr}_{(AB)}[(I\otimes B_k) \rho^{AB} (I\otimes B_k)]$ is the
probability.

\begin{figure}
\begin{center}
{\includegraphics[width=9cm,height=3cm]{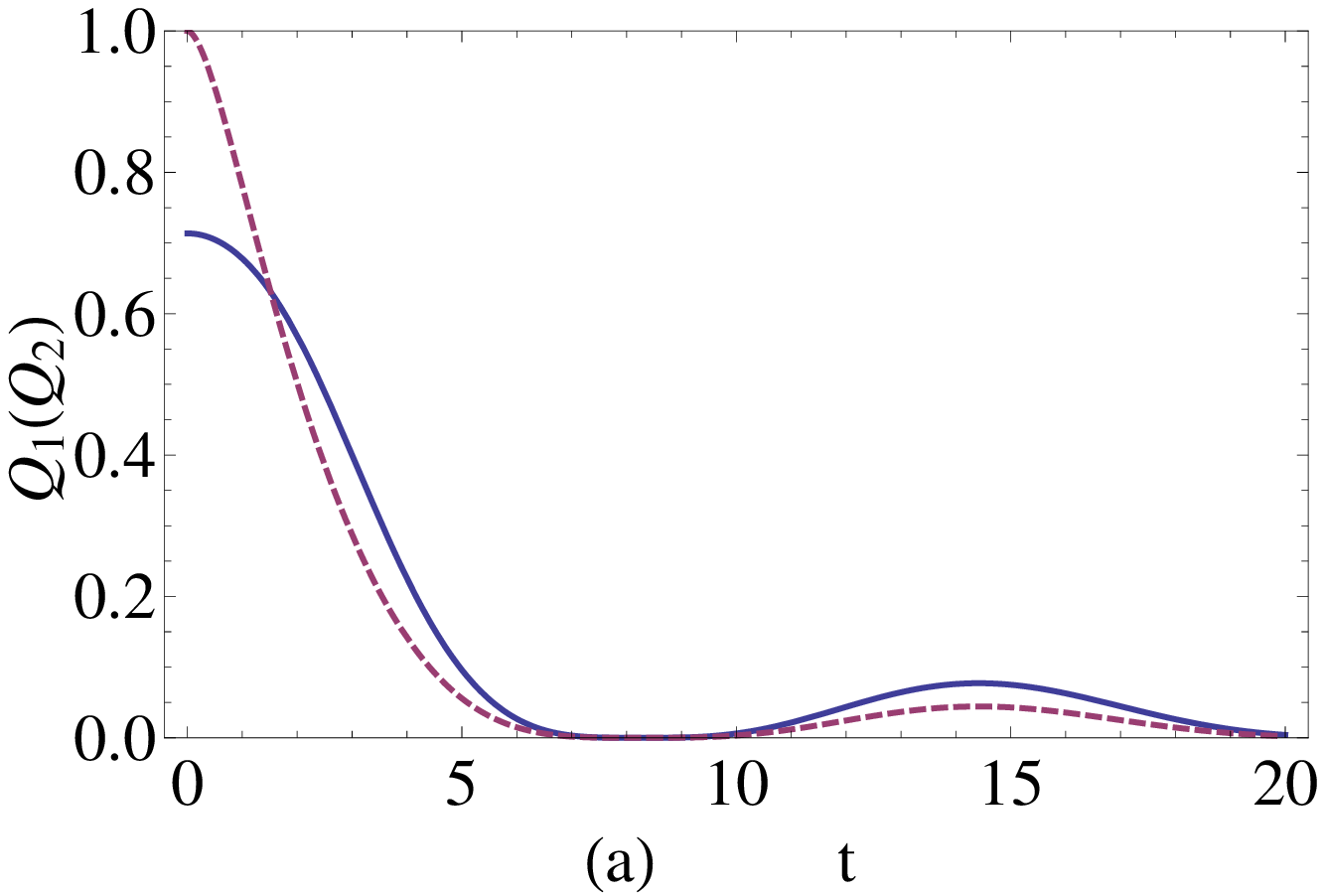}}
{\includegraphics[width=9cm,height=3cm]{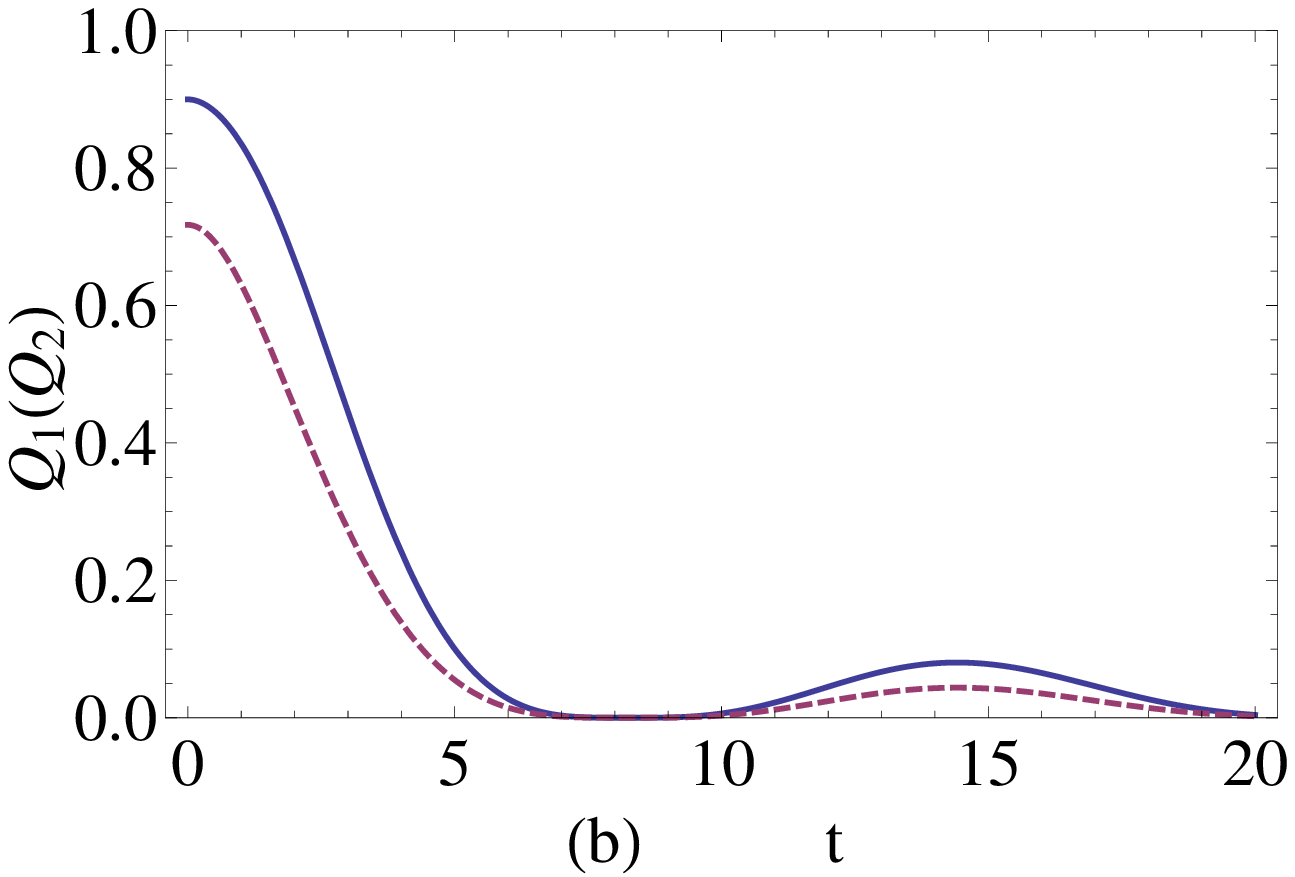}}
\end{center}
\caption{$Q_1(t)$(solid line) and $Q_2(t)$(dash line) are plotted as a function of time $t$ with $\gamma_0=1,\lambda=0.1$
for different initial states (a)$c_1=0.9, c_2=-0.9, c_3=1$ and (b)$c_1=0.9, c_2=-0.9, c_3=0.8$ without control pulses.
The analytical solution of the quantum discord is the minimum value assumed by the functions $Q_1(t)$ and $Q_2(t)$.}
\end{figure}

We assume that the two qubits are initially in the
states with maximally mixed marginals [14], which can be described by the three-parameter
$X$-type density matrix
\begin{eqnarray}
\rho^{AB}(0)=\frac{1}{4}
\left(
\begin{array}{c}
1+c_3 \quad 0 \quad 0 \quad c_1-c_2\\
0 \quad 1-c_3 \quad c_1+c_2 \quad 0\\
0 \quad c_1+c_2 \quad 1-c_3 \quad 0\\
c_1-c_2 \quad 0 \quad 0 \quad 1+c_3
\end{array}\right)\;,
\end{eqnarray}
where $c_i(0\leq c_i\leq 1)$ are the real numbers. The state in Eq. (21) represents a considerable class of states
including the Werner $(|c_1|=|c_2|=|c_3|=c)$ and Bell $(|c_1|=|c_2|=|c_3|=1)$ basis states. Then, we obtain
the density matrix $\rho^{AB}(t)$ for the two noninteracting qubits as
\begin{eqnarray}
\rho^{AB}(t)=
\left(
\begin{array}{c}
a \quad 0 \quad 0 \quad w\\
0 \quad b \quad z \quad 0\\
0 \quad z \quad b \quad 0\\
w \quad 0 \quad 0 \quad d
\end{array}\right)\;,
\end{eqnarray}
where
$$a=\frac{1}{4}(1+c_3)P_t^2, \quad b=\frac{1}{2}P_t-\frac{1}{4}(1+c_3)P_t^2,$$
$$d=1-P_t+\frac{1}{4}(1+c_3)P_t^2,$$
\begin{equation}
z=\frac{1}{4}(c_1+c_2)P_t,\quad w=\frac{1}{4}(c_1-c_2)P_t.
\end{equation}
Using the density matrix $\rho^{AB}(t)$, we obtain the explicit expression of quantum discord between two qubits as [7]
\begin{equation}
Q(\rho^{AB})=\min\{Q_1,Q_2\},
\end{equation}
where
$$
Q_1=S(\rho^{A})-S(\rho^{AB})-a\log_2(\frac{a}{a+b})-b\log_2(\frac{b}{a+b})
$$
\begin{equation}
-d\log_2(\frac{d}{d+b})-b\log_2(\frac{b}{d+b})
\end{equation}
and
\begin{equation}
Q_2=S(\rho^{A})-S(\rho^{AB})-\triangle_+\log_2\triangle_+-\triangle_-log_2\triangle_-,
\end{equation}
with $\triangle_\pm=\frac{1}{2}\pm\frac{1}{2}\sqrt{(a-d)^2+4(|z|+|w|)^2}.$

If no control pulses is presented, $P_t=e^{-\lambda t}[\cos\frac{dt}{2}+\frac{\lambda}{d}\sin\frac{dt}{2}]^2$, which has
discrete zeros at $t_n=2[n\pi-\arctan (d/\lambda)]/d$, with $n$ integer. It is easy to check that $Q(t_n)=Q_1(t_n)=Q_2(t_n)=0$.
For the time near these discrete zeros, for example at time $t_0^-$, we can obtain $Q_1(t_0^-)>Q_2(t_0^-)$ by means of Taylor
series expansion. For the initial state of Eq. (21), we have
$$
Q_1(0)-Q_2(0)=\sum_{k=1}^2\frac{1+(-1)^k|c_3|}{2}\log_2[1+(-1)^k|c_3|]
$$
\begin{equation}
-\sum_{k=1}^2\frac{1+(-1)^k\chi}{2}\log_2[1+(-1)^k\chi],
\end{equation}
where $\chi=\max(|c_1|,|c_2|)$. We can check that $Q_1(0)<Q_2(0)$ if $|c_3|>\max(|c_1|,|c_2|)$ and $Q_1(0)\geq Q_2(0)$ if
$|c_3|\leq\max(|c_1|,|c_2|)$. If $|c_3|>\max(|c_1|,|c_2|)$, noting that both $Q_1(t)$ and $Q_2(t)$ decrease monotonically when $t<t_0$
as Fig. 1(a) shows, there is a point of intersection of $Q_1(t)$ and $Q_2(t)$ when $t$ is between $0$ and $t_0$, i.e., a peculiar dynamics of
quantum correlation $Q(t)$ with a sudden change occurs. However, no sudden change is observed if $|c_3|\leq \max(|c_1|,|c_2|)$, as Fig. 1(b) shows. It should be noticed that, in this paper, the initial parameters for the phenomenon of sudden change occur are different from Ref. [14] because of the different environment. And this phenomenon is not found in Ref. [7] where the quantum discord dynamics of two qubits in independent non-Markovian reservoirs are investigated, but different initial states are chosen there.

\begin{figure}
\begin{center}
{\includegraphics[width=9cm,height=3cm]{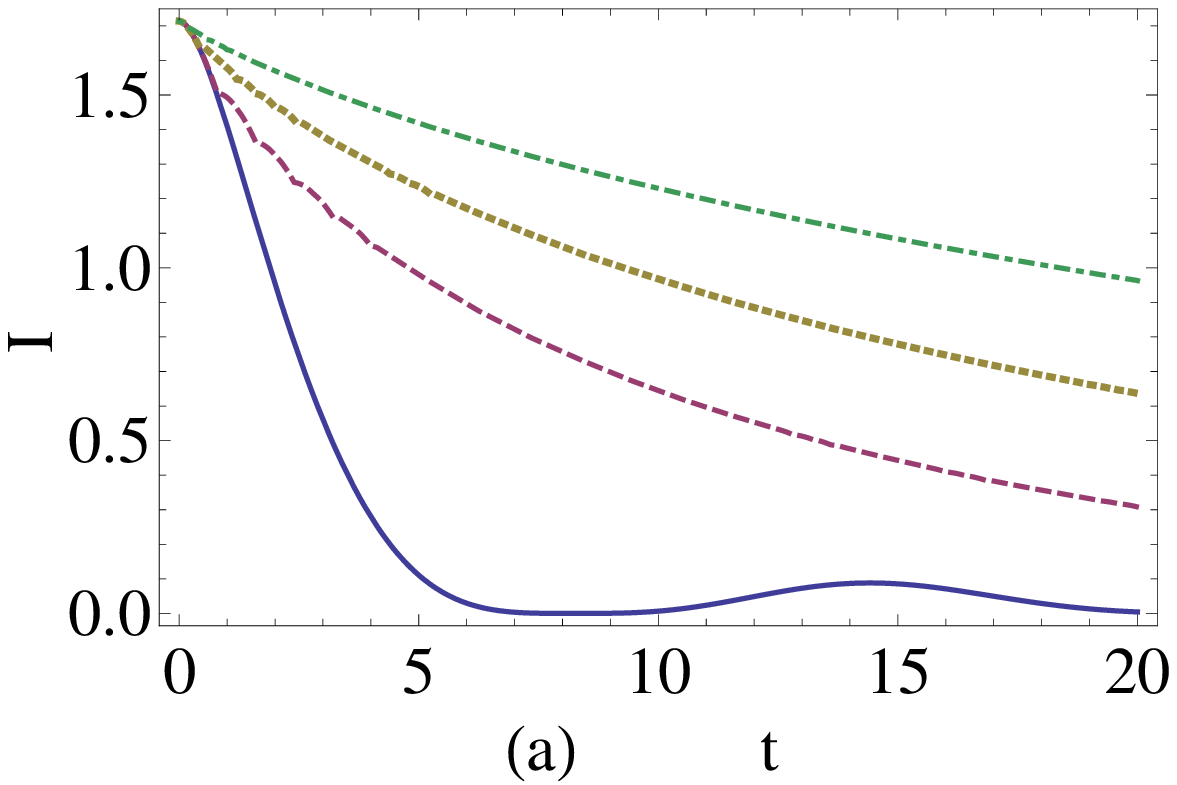}}
{\includegraphics[width=9cm,height=3cm]{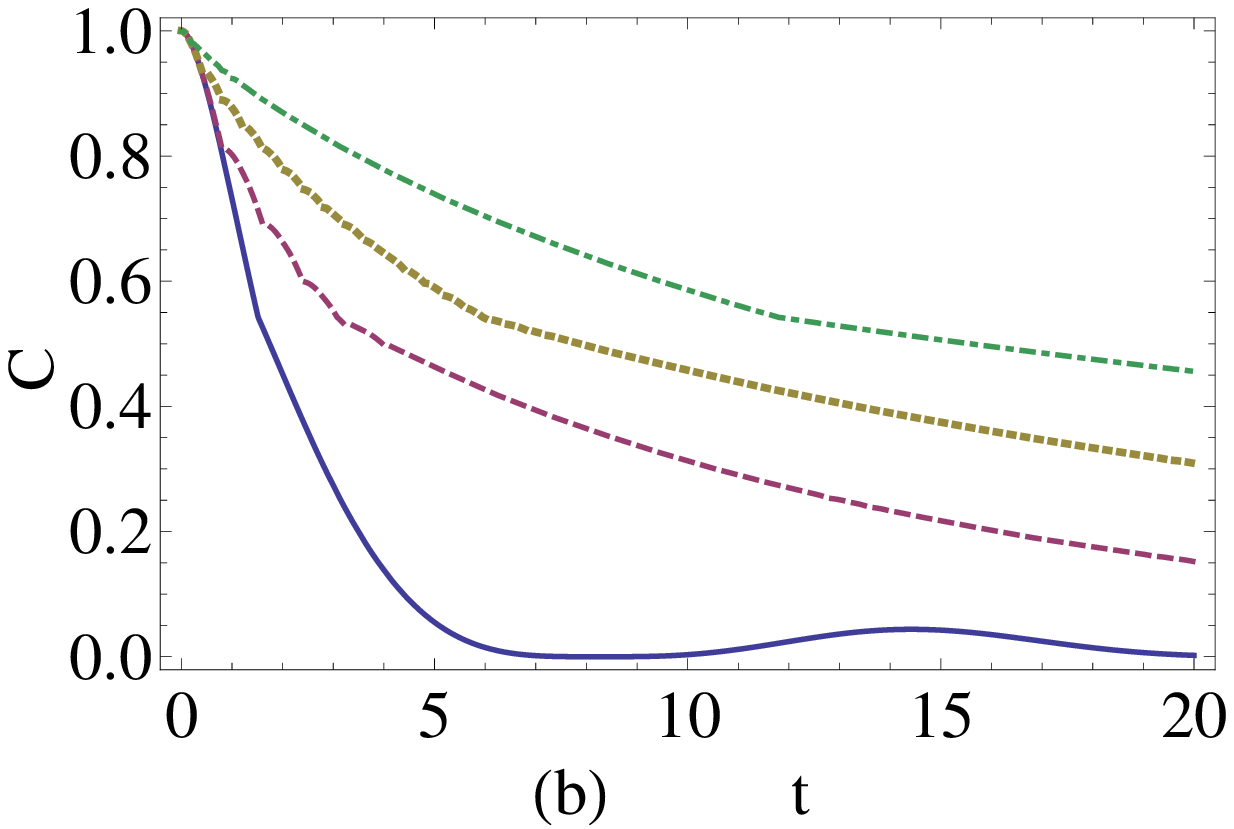}}
{\includegraphics[width=9cm,height=3cm]{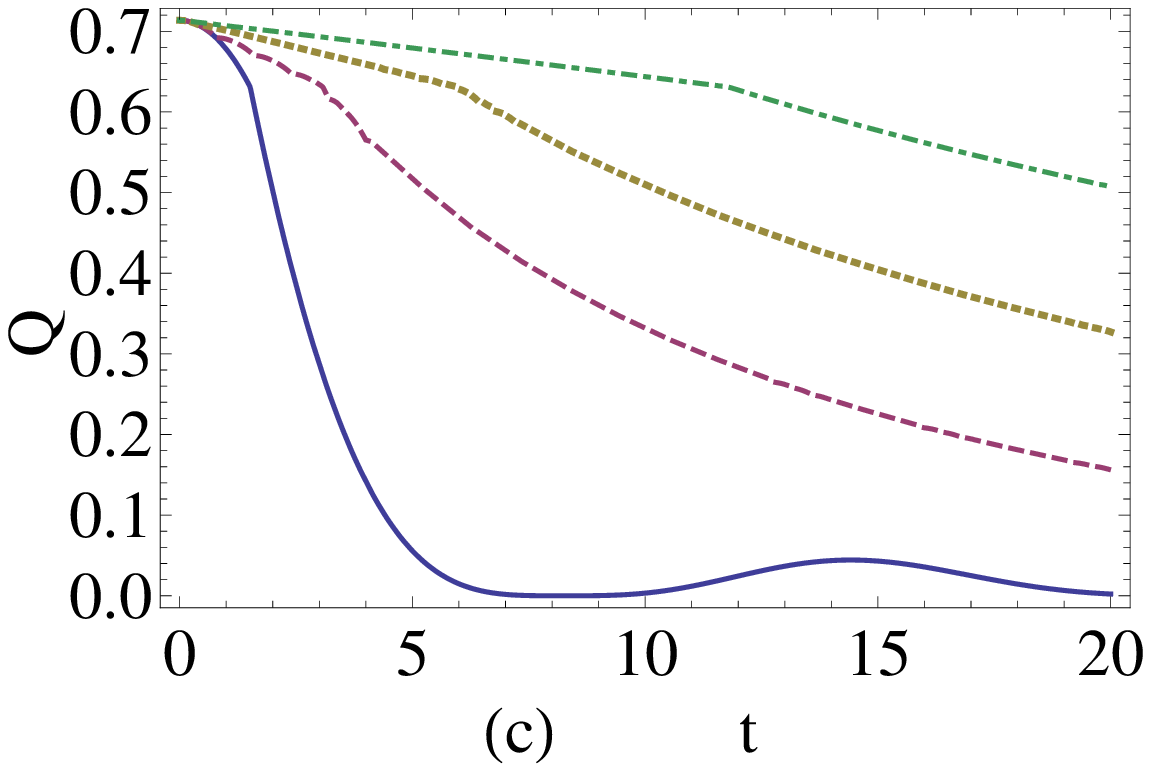}}
\end{center}
\caption{Quantum mutual information $I(t)$(a), classical correlation $C(t)$(b) and Quantum discord $Q(t)$(c)
are plotted as a function of time $t$  with $\gamma_0=1,\lambda=0.1, c_1=0.9, c_2=-0.9, c_3=1$ for different $T$: $T=0.4$(dash line),
$T=0.2$(dot line), $T=0.1$(dot-dash line) and without control pulses(solid line).}
\end{figure}

\begin{figure}
\begin{center}
{\includegraphics[width=6.5cm]{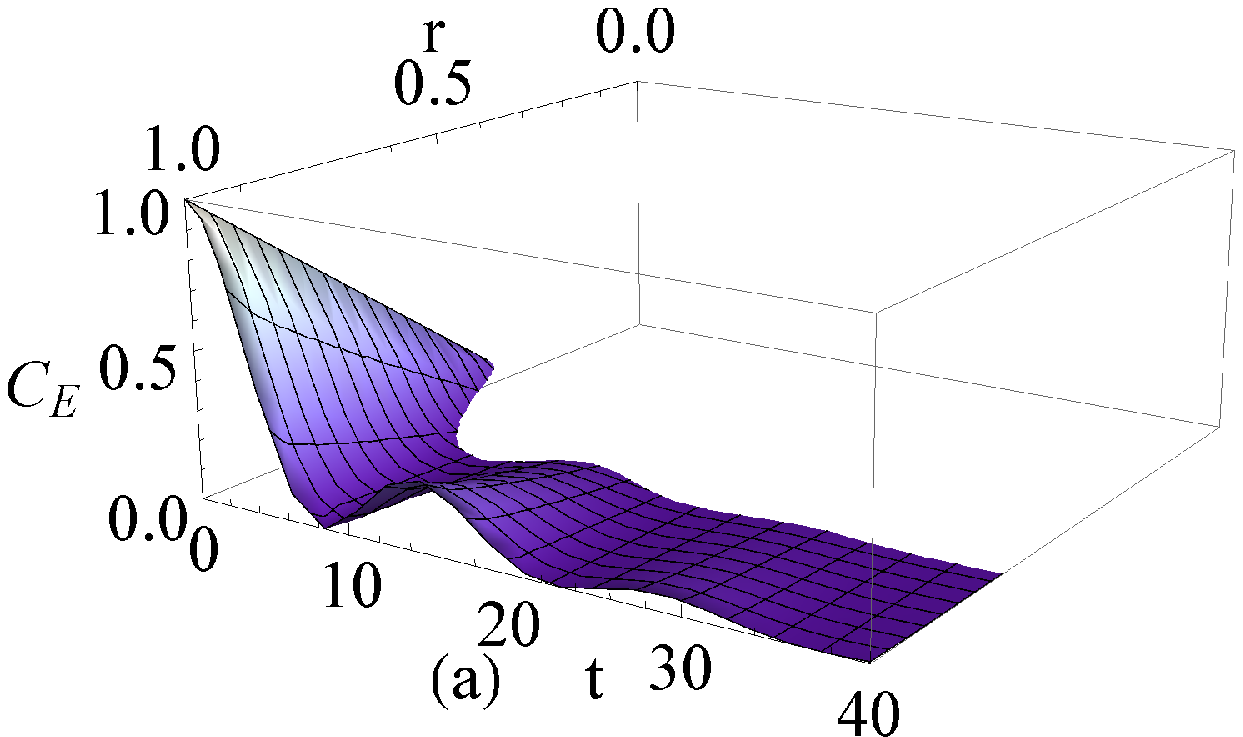}}
{\includegraphics[width=6.5cm]{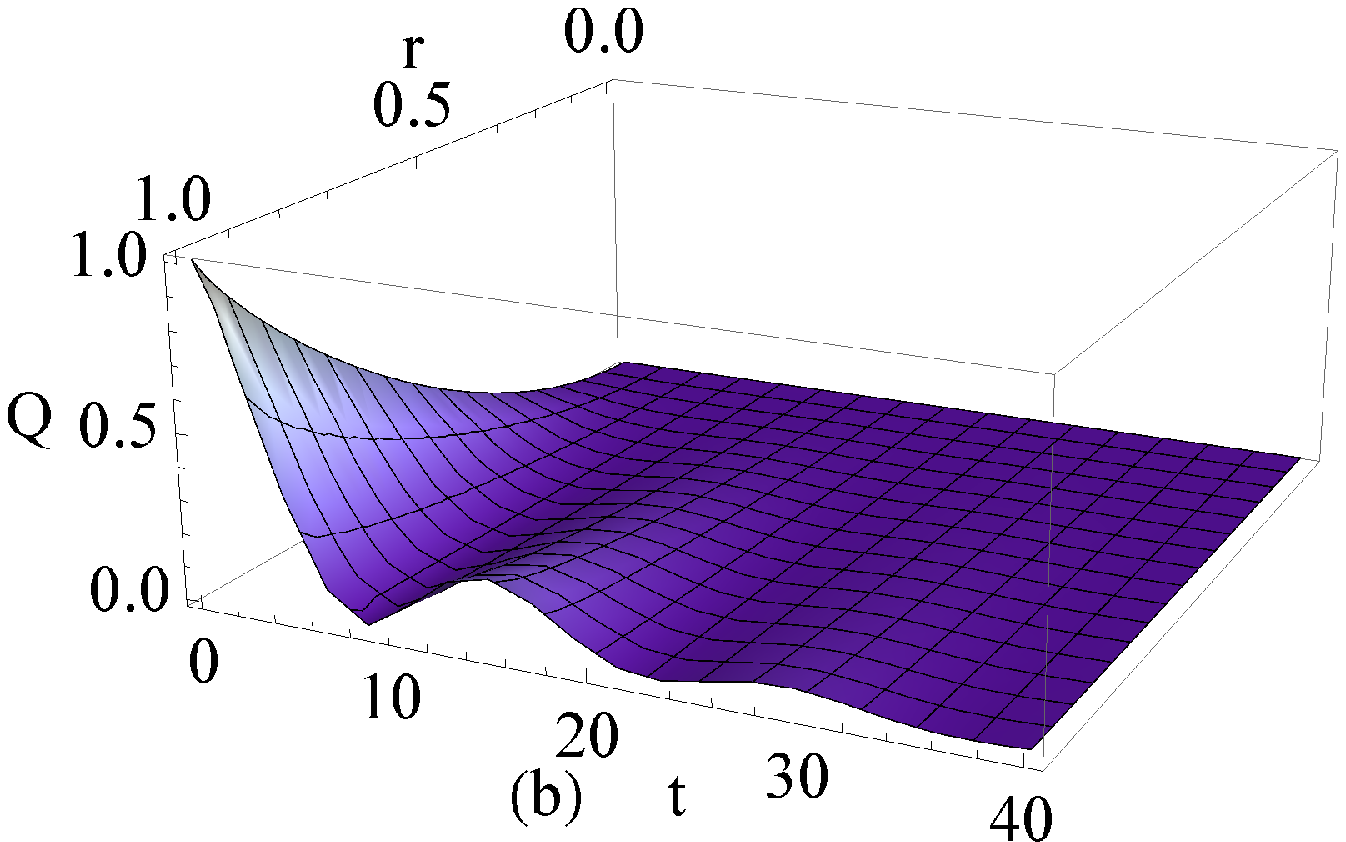}}
\end{center}
\caption{The concurrence $C_E(t)$(a) and the quantum discord $Q(t)$(b) of two atoms are
plotted as a function of $t$ and $r$ for the parameters $\gamma_0=1,\lambda=0.1$ without control pulses.}
\end{figure}

\begin{figure}
\begin{center}
{\includegraphics[width=9cm,height=3cm]{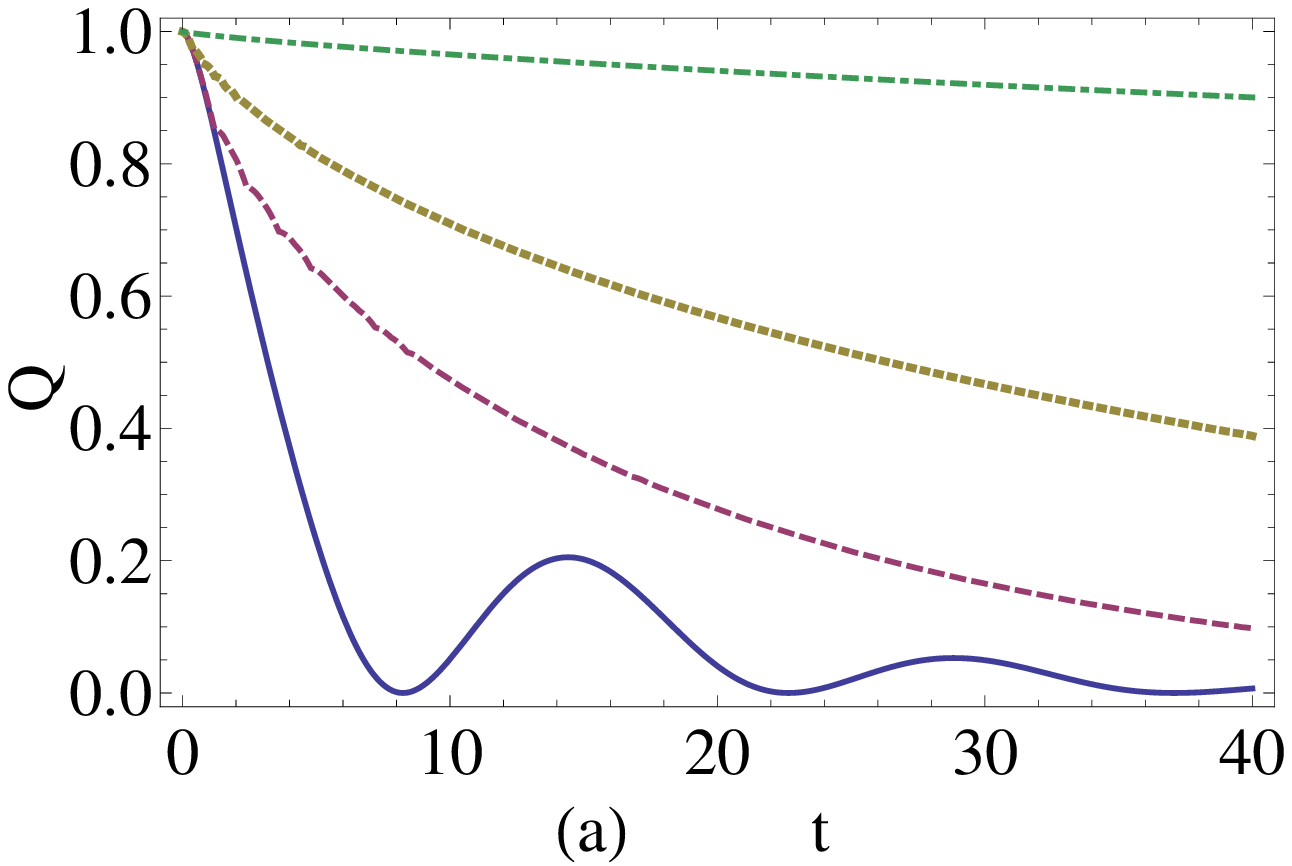}}
{\includegraphics[width=9cm,height=3cm]{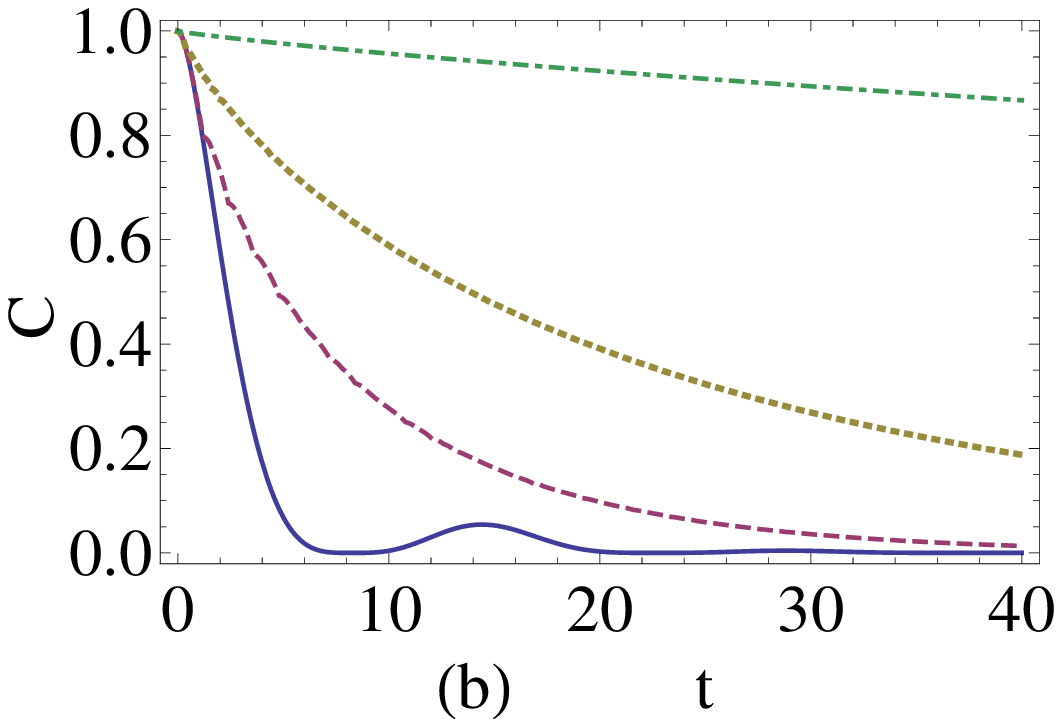}}
\end{center}
\caption{ Quantum discord $Q(t)$(a) and classical correlation $C(t)$(b)
are plotted as a function of time $t$  with $\gamma_0=1,\lambda=0.1, r=1$ for different $T$: $T=0.6$(dash line),
$T=0.2$(dot line), $T=0.01$(dot-dash line) and without control pulses(solid line).}
\end{figure}

\begin{figure}
\begin{center}
{\includegraphics[width=9cm,height=3cm]{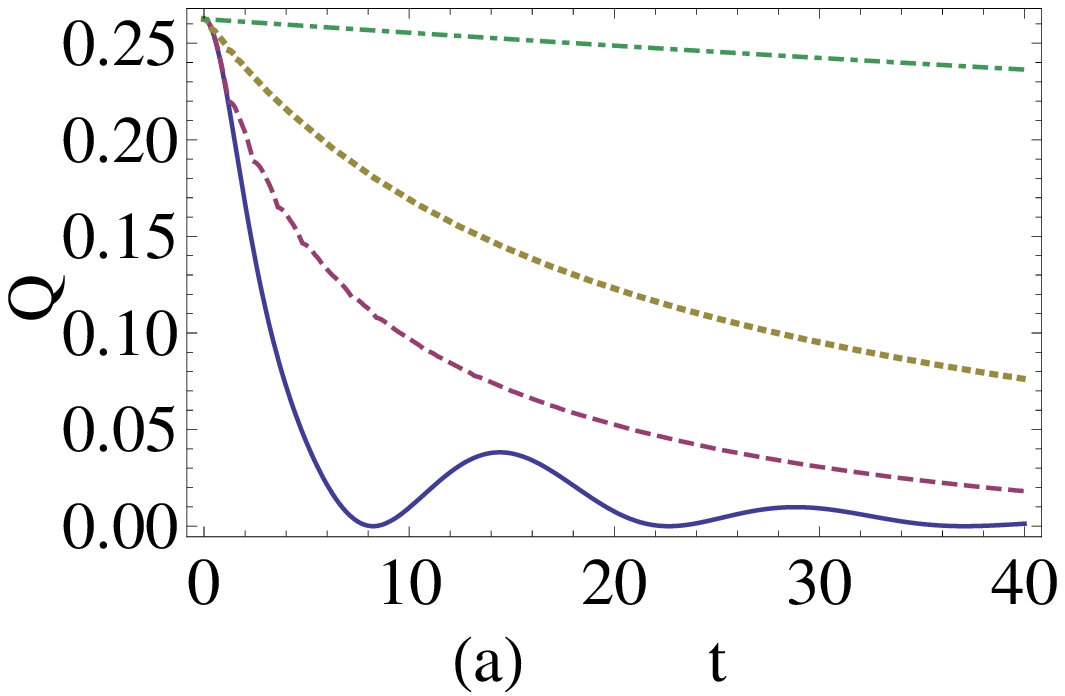}}
{\includegraphics[width=9cm,height=3cm]{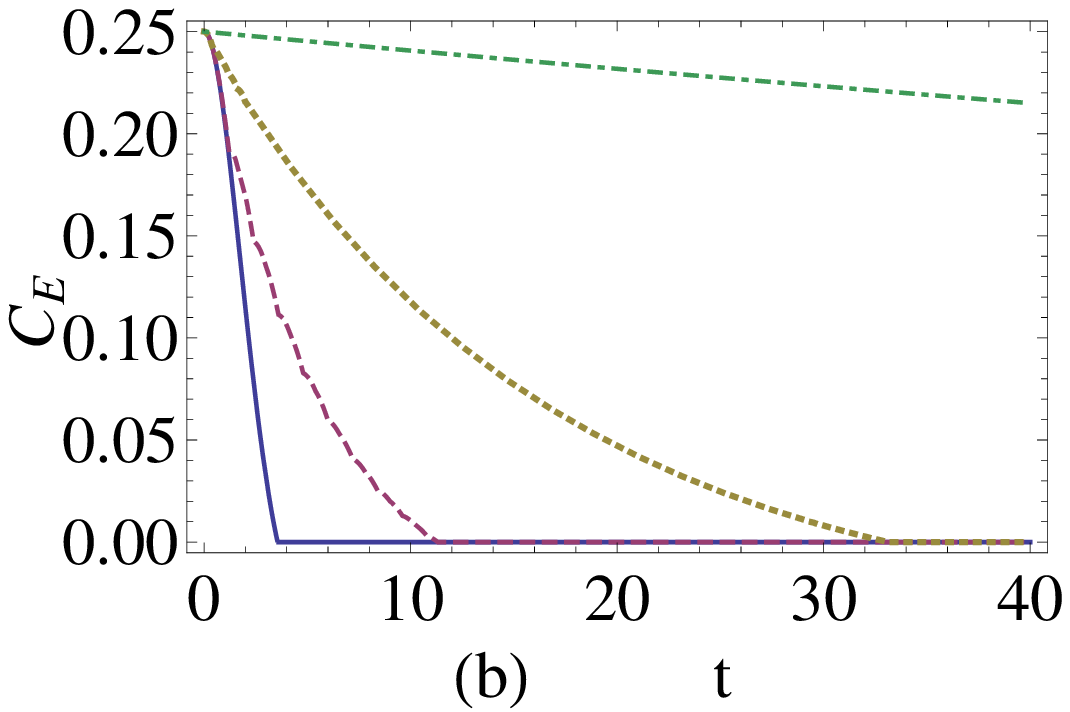}}
\end{center}
\caption{Quantum discord $Q(t)$(a) and concurrence $C_E(t)$(b)
are plotted as a function of time $t$ with $\gamma_0=1,\lambda=0.1, r=0.5$ for different $T$:
$T=0.6$(dash line), $T=0.2$(dot line), $T=0.01$(dot-dash line) and without control pulses(solid line).}
\end{figure}
When the bang-bang control pulses field is presented, quantum mutual information, classical correlation and quantum discord are displayed as a function
of the time $t$ in Fig. 2. It is shown that the bang-bang pulses can protect both the quantum and classical correlations between the two qubits. From Fig. 2(c), we can see that the time when sudden change occurs is prolonged by the bang-bang pulses.

In order to quantify the entanglement dynamics of the qubits, we use Wootters concurrence $C_E(t)$ [18] as a measure. For the initial state of Eq. (21), the explicit expression of the concurrence of two qubits can be obtained from the density matrix $\rho^{AB}(t)$,
\begin{equation}
C_E(t)=max\{0,2|w|-2|b|,2|z|-2\sqrt{ad}\}.
\end{equation}

Now, we choose $c_1=c_2=c_3=-r$, i.e., the initial states of the two noninteracting qubits are the Werner state
\begin{equation}
\rho_W=r|\psi^-\rangle\langle\psi^-|+\frac{1-r}{4}I,
\end{equation}
where $|\psi^-\rangle=(|0\rangle_A|1\rangle_B-|1\rangle_A|0\rangle_B)/\sqrt{2}$ is a maximally entangled state and
$0\leq r\leq 1$. We plot the concurrence and quantum discord as a function of $t$ and $r$ in the absence of control pulses field in Fig. 3.
It can be seen from Fig. 3(a) that entanglement of two qubits is always zero for $r\leq\frac{1}{3}$ [17] and vanishes and revives
with time as $r>\frac{1}{3}$, which means that the entanglement sudden death(ESD) phenomenon [16] appears for the system. However
the quantum discord of two qubits vanishes asymptotically as shown in Fig. 3(b). Different from Ref [12], here, both entanglement and quantum discord cannot recover to the initial values according to the dissipation induced by multi-modes environment. However, they can also be protected effectively by the bang-bang pulses with the same form.

In Fig. 4. the quantum discord and concurrence as a function of $t$ for different values of control pulses are displayed with the parameter $r=1$, i.e., the two atoms are initially prepared in the maximally entangled state. We can see clearly that both the quantum and classical correlations between two qubits can be enhanced by the pulses with short time interval and the smaller interval between the pulses, the more enhancement of the quantum and classical correlations. For the Werner state with $r=0.5$, the concurrence and quantum discord are plotted in Fig. 5 as a function of time $t$ for different intervals of control pulses. We find that both the concurrence and quantum discord can be protected by applying the bang-bang control pulses. It is interesting to point out that the time when ESD occurs can be prolonged by the control pulses.

\begin{figure}
\begin{center}
{\includegraphics[width=7cm]{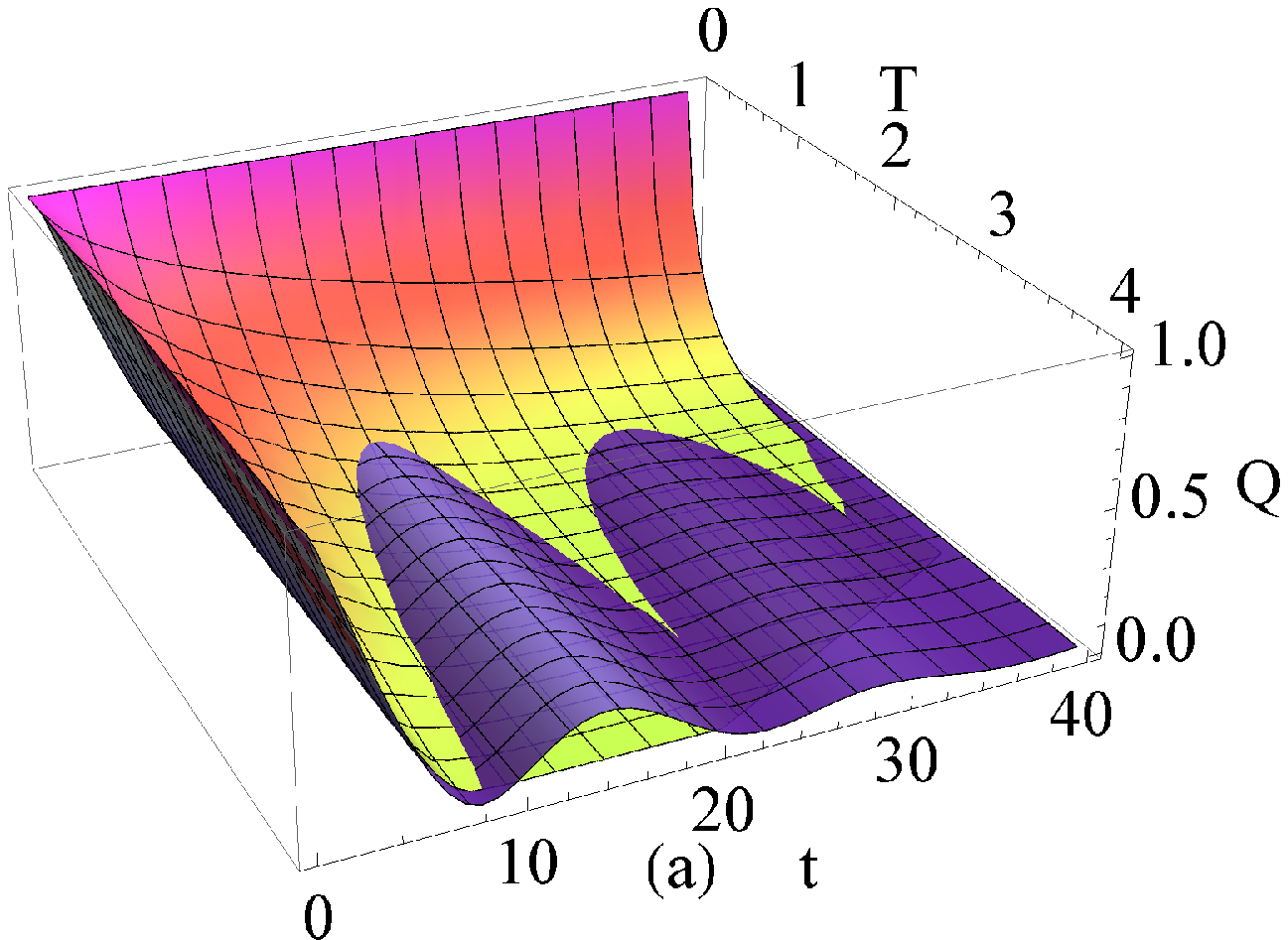}}
{\includegraphics[width=7cm]{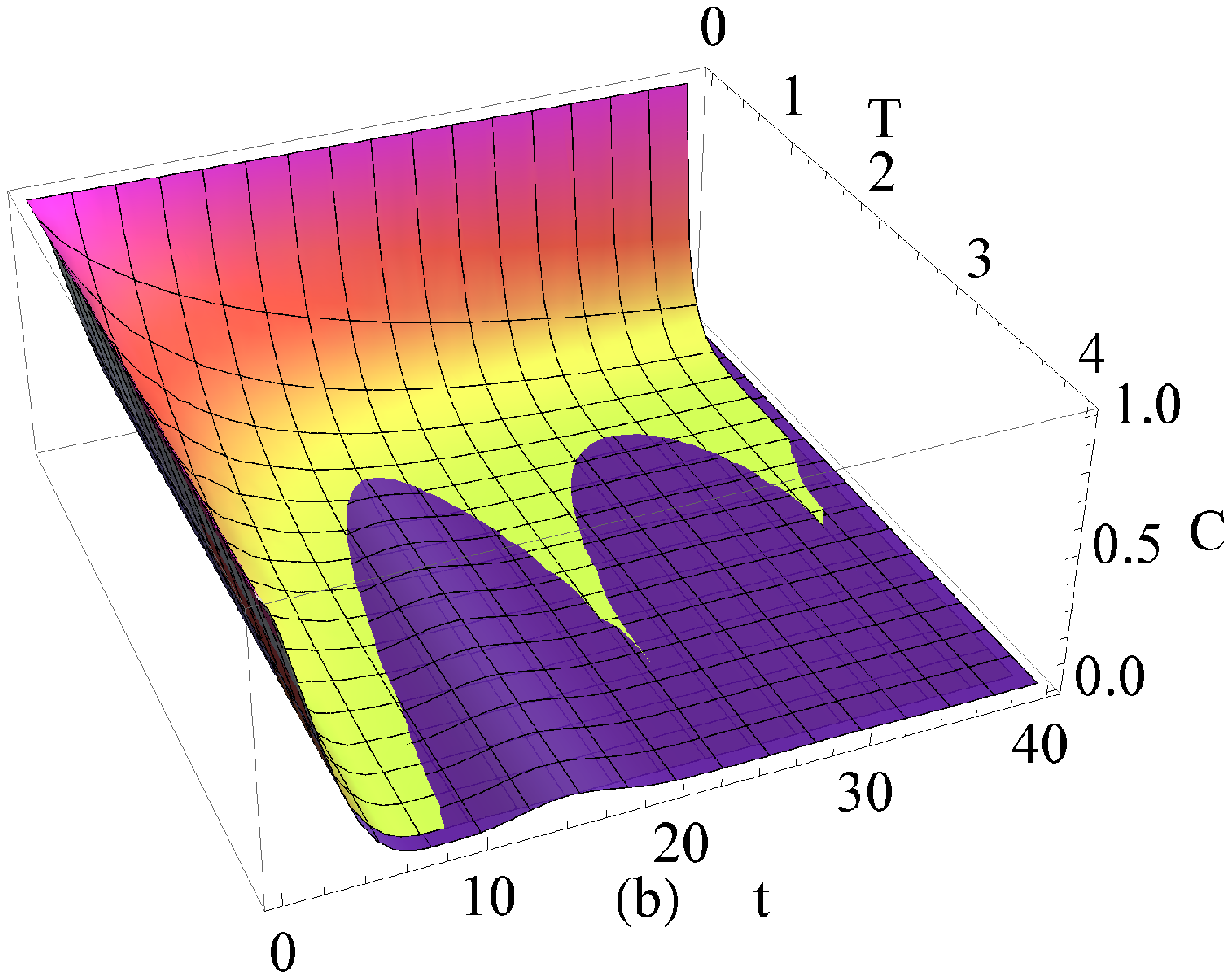}}
\end{center}
\caption{The quantum discord $Q(t)$(a) and classical correlation $C(t)$(b) are plotted as a function of time $t$ and $T$ with $\gamma_0=1,\lambda=0.1, r=1$.
The functions is shown in the present of pulses(NeonColors surface) and in the absence of pulses(T-independent, LakeColors surface). }
\end{figure}

\begin{figure}
\begin{center}
{\includegraphics[width=7cm]{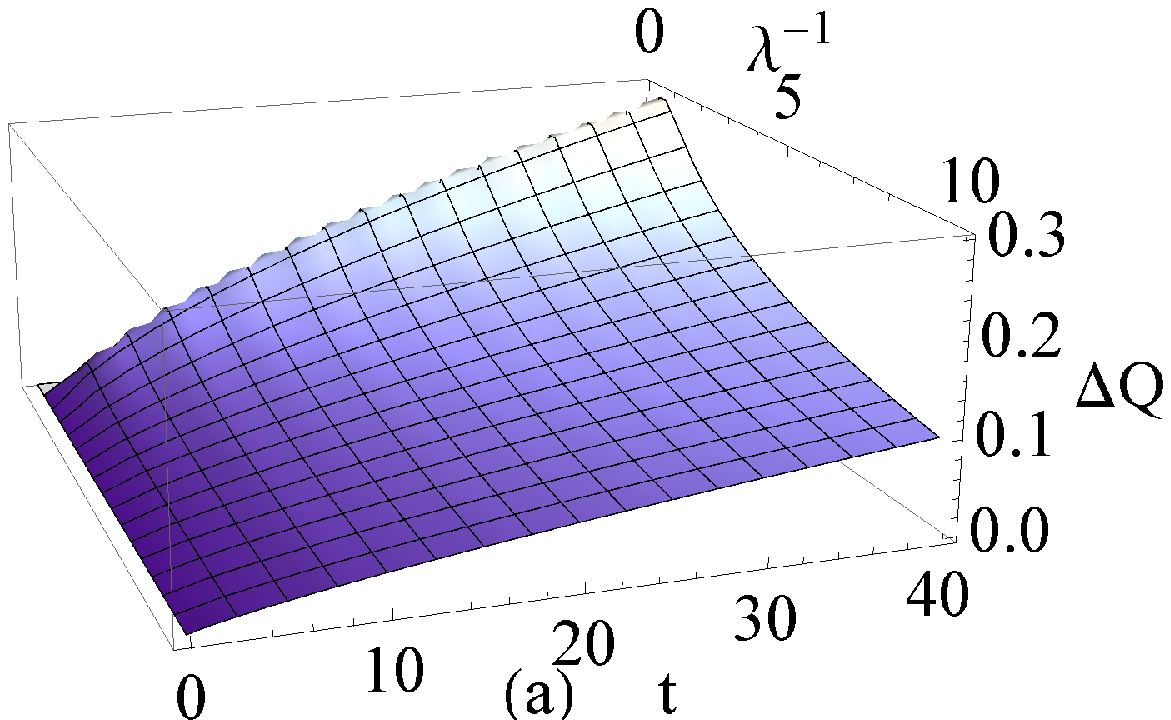}}
{\includegraphics[width=6cm]{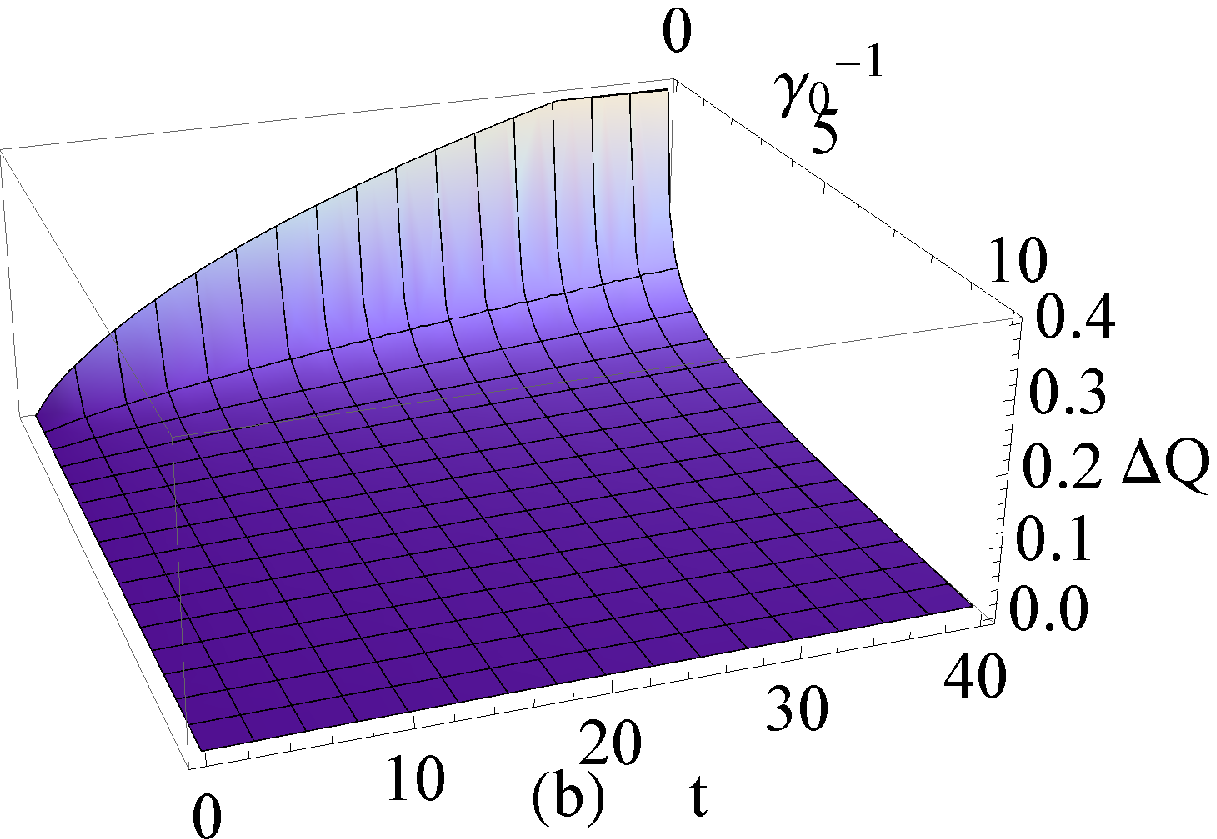}}
\end{center}
\caption{The difference of quantum discord $\Delta Q$ between $Q(0)$ and $Q(t)$ is plotted as a function of time $t$ and (a) $\lambda^{-1}(\simeq \tau_B)$ with $\gamma_0=1, T=0.01, r=1$ or (b) $\gamma_0^{-1}(\simeq \tau_R)$ with $\lambda=0.1, T=0.01, r=1$.}
\end{figure}

We display how the quantum and classical correlations vary with time $t$ and the time interval $T$ in Fig. 6.
It is shown that the pulses with short interval can protect the quantum and classical correlations between two qubits.
Both the quantum and classical correlation descend almost monotonously with time when bang-bang pulses are presented.
When the interval $T$ is long, the value of quantum or classical correlations would be less than the value in the absence of pulses for some region of time $t$. These phenomenon are similar to the quantum Zeno and anti-Zeno effects [19].

In order to explore the influence of the Markovian decay rate and the memory stored in the environment connected to the reservoir correlation time, we display how the difference of quantum discord $\Delta Q(t)=Q(0)-Q(t)$  changes as a function of $\lambda^{-1}(\simeq \tau_B)$ or $\gamma_0^{-1}(\simeq \tau_R)$ in Fig. 7. It is found $\Delta Q$ is smaller, i.e., the quantum correlations are protected more effectively by bang-bang pulses for longer reservoir correlation time $\tau_B$ or relaxation time scale $\tau_R$.

\section{Conclusions}
In this paper, we have proposed a scheme of protecting quantum
correlations for two qubits in independent non-Markovian reservoir by making use of the bang-bang pulses.
Considering the two qubits are initially in the states with maximally mixed marginals, we show that the quantum discord dynamics presents sudden changes for some given choices of initial parameters. We also explore the influence of the bang-bang pulses on the quantum discord and entanglement of two qubits and the phenomenon of sudden change. The quantum correlations between two qubits are protected more effectively with shorter interval pulses or longer reservoir correlation time. Particularly, the time when ESD occurs can be prolonged by the control pulses and the time when sudden change occurs can also be prolonged. We only consider the ideal bang-bang pulses with the duration to be infinitely small and there is no interaction between two qubits. For coupled two qubits, the quantum correlations may be protected by the bang-bang pulses used in this paper. However, protecting the dynamics of coupled quantum systems is a problem needs further consideration since the decoupling pulses may disrupt the interqubit dynamics [20]. It is interesting to investigate the effect of finite duration time of the pulses and other pulse error for more realistic case [21]. The approach adopted here may be used to improve the implementation of quantum information and quantum computation.

\section*{Acknowledgement}
This project was supported by the National Natural Science Foundation of China (Grant No. 10774131).

\end{document}